\documentclass[11pt,twoside,onecolumn]{article}
\usepackage[]{latexsym}
\usepackage{epsfig}
\usepackage{amsmath,amssymb}
\newcommand{\be}{\begin{eqnarray}}
\newcommand{\ee}{\end{eqnarray}}

\title{\bf Black Hole Remnants at the LHC}
\author{
L.~Bellagamba$^{b}$\thanks{lorenzo.bellagamba@bo.infn.it},
$\ $
R.~Casadio$^{a,b}$\thanks{roberto.casadio@bo.infn.it},
$\ $
R.~Di~Sipio$^{a,b}$\thanks{riccardo.disipio@bo.infn.it}
$\ $
and
V.~Viventi$^{a}$\thanks{vi.strawberry@gmail.com}
\\
\null
\\
$^a${Dipartimento di Fisica, Universit\`a di Bologna}
\\
via Irnerio~46, 40126 Bologna, Italy
\\
\\
$^b${Istituto Nazionale di Fisica Nucleare, Sezione di Bologna}
\\
via Irnerio~46, 40126 Bologna, Italy
\\
}
\begin{document}
\maketitle
\begin{abstract}
We investigate possible signatures of black hole events at the LHC
in the hypothesis that such objects will not evaporate completely,
but leave a stable remnant.
For the purpose of defining a reference scenario, we have employed the publicly
available Monte Carlo generator CHARYBDIS2, in which the remnant's
behavior is mostly determined by kinematic constraints and conservation
of some quantum numbers, such as the baryon number.
Our findings show that electrically neutral remnants are highly favored and
a significantly larger amount of missing transverse momentum is to be expected 
with respect to the case of complete decay.
\end{abstract}
\setcounter{page}{1}
\section{Introduction}
\setcounter{equation}{0}
Models with large extra dimensions~\cite{arkani,RS} allow for
the fundamental scale of gravity $M_{\rm G}$ to be as low as the electro-weak
scale ($M_{\rm G}\simeq 1\,$TeV), and microscopic
black holes (BH) may therefore be produced in our
accelerators (for reviews, see, {\em e.g.\/}, Refs.~\cite{cavaglia}).
Once a BH is formed, and the balding phase is over,
the Hawking radiation~\cite{hawking} sets off.
The standard description of this effect is based on the
canonical Planckian distribution for the emitted particles, which implies
the life-time of microscopic BHs is very short, of the order of
$10^{-26}\,$s~\cite{dimopoulos}.
This picture (restricted to the ADD scenario~\cite{arkani}) has been implemented in several
numerical codes~\cite{charybdis,trunoir,ahn,catfish,cha2,blackmax,charybdis2},
which have been mainly designed to extract information that will allow us to identify
BH events at the Large Hadron Collider (LHC)
(see Ref.~\cite{CMS} for a search of BH signatures with the first data collected by the
LHC and Ref.~\cite{park} for some theoretical criticisms on the validity of that
analysis).
\par
It is important to recall that the end-stage of the BH evaporation remains an open issue
(see, {\em e.g.}, Refs.~\cite{Landsberg:2006mm,Harris:2004xt,Casanova:2005id}),
because we do not yet have a confirmed theory of quantum gravity.
In fact, the semiclassical Hawking temperature grows without bound,
as the BH mass decreases, which can be viewed as a sign of the lack of predictability
of perturbative approaches~\footnote{For reviews of the more consistent
microcanonical description of BH evaporation in which energy
conservation is granted, see Refs.~\cite{mfd,entropy}.}.
This is an important issue also on a purely experimental side, since
deviations from the Hawking law for small BH mass (near the fundamental
scale $M_{\rm G}$) could actually lead to detectable signatures.
\par
Our ignorance of the late stages of the BH evaporation is usually bypassed
in the numerical codes by letting the microscopic BH
decay into a few standard model particles as an (arbitrary) lower mass
is reached, but the possibility of ending the evaporation by leaving a
stable remnant has also been considered~\cite{Koch:2005ks,Hossenfelder:2005bd,gingrich}.
The purpose of this report is precisely to study the case in which BHs
stop emitting and leave a long-lived (or even stable) remnant of mass
$M\simeq M_{\rm G}$.
\par
For this purpose, we have employed the Monte Carlo code CHARYBDIS2~\cite{charybdis2},
which is presently the only generator which natively supports remnant BHs, as
we shall describe in Section~\ref{sec:MCcode}.
In Section~\ref{sim}, we report results from test runs and describe in detail both the
primary emission and final output obtained using Pythia \cite{pythia}.
We shall use units with $c=\hbar=1$ and the Boltzmann constant $k_{\rm B}=1$.
\section{Black hole production and evolution}
\setcounter{equation}{0}
In the ADD scenario~\cite{arkani}, a microscopic BH with horizon radius
shorter than the typical size of the extra dimensions is usually approximated by a
$(4+d)$-dimensional solution of the vacuum Einstein equations.
In the simplest case of a non-rotating BH, this leads to the following
expression for the horizon radius,
\be
R_{\rm H}=\frac{1}{\sqrt{\pi}\,M_{\rm G}}\,
\left(\frac{M}{M_{\rm G}}\right)^{\frac{1}{d+1}}
\left(\frac{8\,\Gamma\left(\frac{d+3}{2}\right)}{d+2}
\right)^{\frac{1}{d+1}}
\ ,
\ee
where $M$ is the BH mass, $\Gamma$ the Gamma function,
and the temperature associated with the horizon is given by
\be
T_{\rm H} =
\frac{d+1}{4\,\pi\,R_{\rm H}}
\ .
\label{TH}
\ee
\par
At the LHC, a BH could form by colliding two protons.
The total BH cross section can be estimated from the geometrical
hoop conjecture~\cite{thorne} as
\be
\sigma(M)\approx \pi\,R_{\rm H}
\ .
\ee
In order to determine the total production cross section, this expression
must be rescaled according to the parton luminosity approach as
\be
\left.\frac{d\sigma}{d M}\right|_{pp\to BH+X}
=\frac{dL}{d M}\,\hat\sigma(ab\to BH; \hat s=M^2)
\ ,
\ee
where $a$ and $b$ represent the partons which form the BH,
$\sqrt{\hat s}$ is their centre-mass energy and
\be
\frac{dL}{d M}=\frac{2\,M}{s}\,\sum_{a,b}\int_{M^2/s}^1
\frac{dx_a}{x_a}\,f_a(x_a)\,f_b\left(\frac{M^2}{s\,x_a}\right)
\ ,
\ee
where $f_i(x_i)$ are parton distribution functions (PDF)
and $\sqrt{s}$ the LHC centre-mass collision energy
(up to $7\,$TeV presently, with a planned maximum of $14\,$TeV).
Of course, semiclassical BHs will only form above
a minimum mass (presumably) larger than
$M_{\rm G}$~\cite{park,scale}.
\par
Once formed, the BH begins to evolve.
In the standard picture, the decay process can be divided
into three characteristic stages:
\begin{description}
\item[\em Balding phase:]
the BH radiates away the multipole moments inherited from the initial
configuration and reaches a hairless state.
A fraction of the initial mass will also be radiated as gravitational radiation.
\item[\em Evaporation phase:]
the BH loses mass via the Hawking effect.
It first spins down by emitting the initial angular momentum,
after which it proceeds with the emission of thermally distributed
quanta.
The radiation spectrum contains all the Standard Model (SM) particles,
(emitted on our brane), as well as gravitons (also emitted into the
extra~dimensions).
For this tags, it is crucial to have a good estimate of the grey-body
factors~\cite{ida,kanti}.
\item[\em Planck phase:]
the BH has reached a mass close to the effective Planck
scale $M_{\rm G}$ and falls into the regime of quantum gravity.
It is generally assumed that the BH will either completely
decay into SM particles~\cite{dimopoulos} or a (meta-)stable
remnant is left, which carries away the remaining energy~\cite{Koch:2005ks}.
\end{description}
We will here focus on the possibility that the third phase
ends by leaving a stable remnant.
\section{Remnants and a Monte Carlo code}
\label{sec:MCcode}
\setcounter{equation}{0}
Several Monte Carlo codes which simulate the production and decay
of micro-BHs are now available  (see, {\em
e.g.}~Refs.~\cite{charybdis,trunoir,ahn,catfish,cha2,blackmax,charybdis2}).
The only code which presently implements the description of a remnant
is CHARYBDIS2~\cite{charybdis2}~\footnote{This code has been used to
study the remnants of non-commutative BHs in Ref.~\cite{gingrich}}.
Such a description is mostly based on kinematic and quantum number constraints,
so that the remnant will necessarily have zero baryon number.
The production and evaporation phases are better modeled, from a
theoretical point of view, although they could still be improved.
For example, the production phase is described using bounds from trapped
surface formation, which might lead to overestimate the amount of missing
energy in gravitational radiation~\cite{sampaio1}.
Further refinements have also been investigated for the Hawking
spectra~\cite{sampaio2,cha2}.  
\par
A list of the relevant adjustable parameters in the code and the values we
used in the simulations, is given in Table~\ref{FreePar}.
In the study we present here, we simulated the different scenarios in $pp$
collisions with total centre-mass energy of both
$\sqrt{s}=7\,$TeV~\footnote{This case was mostly included for a comparison
with the CMS analysis in Ref.~\cite{CMS},
although there are arguments which show that $7\,$TeV might not be
theoretically enough to produce semiclassical BHs~\cite{park}.}
and $14\,$TeV.
\begin{table}[t]
\centerline{
\begin{tabular}{|c|c|c|}
\hline
Parameter & Description & Used values
\\
\hline
$M_{\rm G}$ & Fundamental gravitational scale & $2-4$ TeV
\\
\hline
D & Total number of dimensions & 6, 8
\\
\hline
MINMSS & minimum initial BH mass & $2 \times M_{\rm G}$ 
\\
\hline
MAXMSS & maximum initial BH mass & $\sqrt{s}$ 
\\
\hline
RMSTAB & main switch for stable BH remnant & TRUE/FALSE
\\
\hline
KINCUT & kinematic condition for BH remnant & TRUE/FALSE
\\
\hline
\end{tabular}
}
\caption{Most relevant parameters of CHARYBDIS2 and their values
for the different simulated samples used in this study.}
\label{FreePar}
\end{table}
\par
\begin{figure}[t]
\centerline{
\begin{tabular}{ r}
\centerline{\hbox{
\epsfig{figure=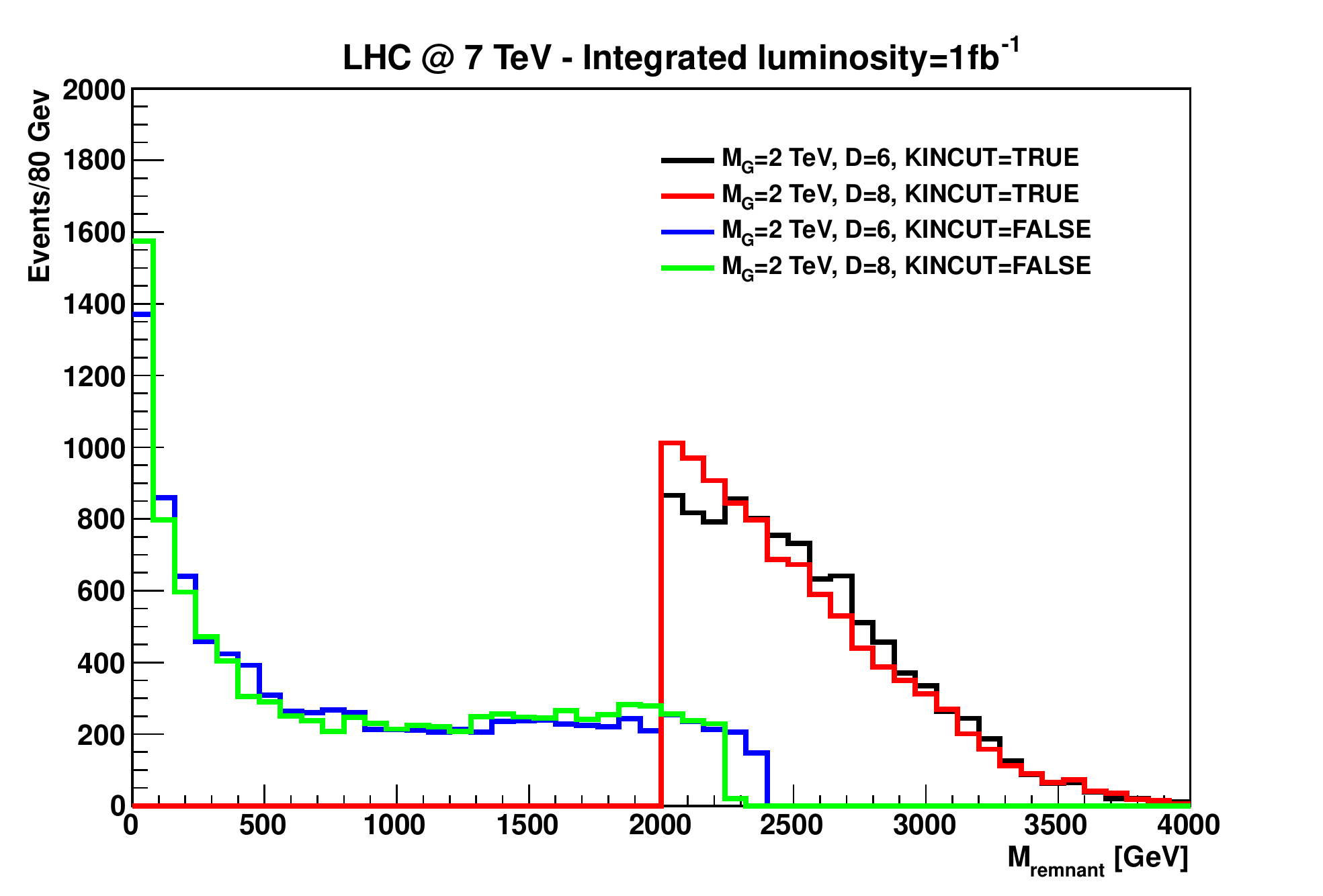,height=10cm,clip=} }}
\end{tabular}}
\caption{Remnant mass for different values of the KINCUT parameters.}
\label{fig:remmas}
\end{figure}

It is particularly important to describe in details the role of one of the
kinematic constraints, namely KINCUT.
When set to TRUE, the generator will not allow the emission of Hawking
quanta that reduce the BH mass $M$ below $M_{\rm G}$,
whereas KINCUT=FALSE will allow the last decay particle to have
an energy as large as the BH's (but no more).
Let us then see what will likely happen for the last emission,
when the BH has already evaporated down to a mass
$M_{\rm LE}\gtrsim M_{\rm G}$.
The generator will try to simulate an Hawking particle with energy
$\omega$ determined by the Planckian distribution with a
temperature $T_{\rm H}=T_{\rm H}(M_{\rm LE})$
and small enough to ensure energy conservation.
Roughly speaking (that is, neglecting the BH kinetic energy),
this means $T_{\rm H}(M_{\rm LE})\simeq\omega<M_{\rm LE}$.
Since the temperature~\eqref{TH} increases for decreasing $M$,
it is likely the attempted last $\omega$ will carry away enough
energy to reduce $M$ below $M_{\rm G}$.
If KINCUT=TRUE, this process is forbidden and the generator simply
stops and leaves a remnant of mass $M_{\rm LE}>M_{\rm G}$.
If KINCUT=FALSE, the process is allowed and the remnant mass
will be $M\simeq M_{\rm LE}-\omega<M_{\rm G}$.
The effect of KINCUT is clearly visible in Fig.~\ref{fig:remmas},
which shows the remnant mass distributions for samples generated
at $\sqrt{s}=7\,$TeV with $M_{\rm G}=2\,$TeV, in $D=6$ and $8$.
If KINCUT=FALSE the last decay step, which allows the remnant to access
masses lower than $M_{\rm G}$, typically releases a lot of energy,
producing low mass remnants. 
On the other hand KINCUT=TRUE forbids remnant masses below $M_{\rm G}$, 
thus producing an increase of events as the remnant mass approaches 
$M_{\rm G}$ and long tails towards higher values. 
These scenarios are in agreement with the results we shall show
in the next Section.
\par
From a theoretical point of view, one expects a remnant mass
around $M_{\rm G}$ (see, {\em e.g.},~Ref. \cite{scale}).
The above scenarios are therefore, in a sense, complementary,
with the KINCUT=FALSE case perhaps less sound.
\section{Simulations}
\setcounter{equation}{0}
\label{sim}
The standard setting for CHARYBDIS2 simulates the evaporation of a
microscopic BH according to Hawking's law, until the black
hole mass $M$ reaches the minimum value, which is assumed equal
to the fundamental scale of gravity $M_{\rm G}$.
The BH subsequently decays into a few bodies ($N_{\rm f}=2,\ldots,6$)
simply according to phase space.
As described in the previous Section, we are instead interested in the study 
of scenarios with a residual stable BH remnant in the final state. 
\par
In order to compare the output of the different scenarios (no stable remnant,
stable remnant with KINCUT=TRUE/FALSE) we have generated Monte Carlo samples
varying the parameters listed in Table~\ref{FreePar} and assuming two 
different centre mass energy for the LHC, namely $7$ and $14\,$TeV.
As pointed out in Table~\ref{FreePar}, the mass of the produced BH has
been required to be at least $2\,M_{\rm G}$ in order to avoid
the region where the classical theory, used to evaluate the production 
cross section, is no longer valid (but see also Ref.~\cite{park}).
The other parameters have been fixed to the CHARYBDIS2 default values. 
Table~\ref{tab:mcsamples} summarizes the characteristics
of the generated samples, including the $t\bar{t}$ production process
which is one of the most relevant backgrounds.
The table also contains the event yields normalized to an integrated luminosity
of $1\,$fb$^{-1}$  for few basic selection requirements applied at the hadronization
level, as described in details in Section~\ref{sec:results}.
\begin{table}[t]
{\scriptsize
\centerline{
\begin{tabular}{|c|c|c|c|c|c|c|c|} 
\hline
\multicolumn{8}{|c|}{$pp$ collision $\sqrt{s}=7\,$TeV}
\\
\hline
Process & $M_{\rm G}$ [TeV] & D & $\sigma$ [pb] & remnant & 
$P_T^{\rm lep}>0.5\,$TeV & $P_T^{\rm miss}>0.5\,$TeV & $E_T>2\,$TeV  
\\
\hline
 BH & 2.0 & 6 & 0.017 & KINCUT=TRUE  & $0.18\pm0.02$   &   $11.8\pm0.1$   &   $0.39\pm0.03$  
\\
\hline
 BH & 2.0 & 6 &  "  & KINCUT=FALSE & $0.51\pm0.03 $   &  $14.9\pm0.2$    &  $7.1\pm0.1$ 
\\
\hline
 BH & 2.0 & 6 &  "  & no remnant   &  $0.94\pm0.04$  &    $1.74\pm0.05$  &    $15.8\pm0.2$
\\
\hline
 BH & 2.0 & 8 & 0.052 & KINCUT=TRUE & $0.49\pm0.05$   &    $33.9\pm0.4$  &   $1.69\pm0.09$
\\
\hline
 BH & 2.0 & 8 &  "  & KINCUT=FALSE & $1.57\pm0.09$  &    $43.0\pm0.5$   &   $22.9\pm0.3$ 
\\
\hline
 BH & 2.0 & 8 &  "  & no remnant   & $3.2 \pm0.1$   &    $5.4 \pm0.2$   &   $46.5\pm0.5$
\\
\hline
 BH & 2.5 & 8 & $1.74\times 10^{-2}$ & KINCUT=TRUE & $0.0023\pm0.0002$ &  $0.130\pm0.001$ & $0.0169\pm0.0005$
\\
\hline
 BH & 2.5 & 8 &  "  & KINCUT=FALSE & $0.0070\pm0.0003$  & $0.151\pm0.002$ &  $0.112\pm0.001$
\\
\hline
 BH & 2.5 & 8 &  "  & no remnant   & $0.0119\pm0.0005$  & $0.0225\pm0.0006$ & $0.162\pm0.002$
\\
\hline
 $t\bar{t}$ & - & - & 84 & - & $0$ &   $2.0\pm1.2$   & $5.4\pm1.9$
\\
\hline
\hline
\multicolumn{8}{|c|}{$pp$ collision $\sqrt{s}=14\,$TeV}
\\
\hline
Process & $M_{\rm G}$ [TeV] & D & $\sigma$ [pb] & remnant & $P_T^{\rm lep}>1\,$TeV
& $P_T^{\rm miss}>1\,$TeV & $E_T>4\,$TeV  
\\
\hline
 BH & 3.5 & 6 &  0.040  & KINCUT=TRUE  & $0.30\pm0.03$  & $24.8\pm0.3$ & $0.47\pm0.04$
\\
\hline
 BH & 3.5 & 6 &   "  & KINCUT=FALSE  & $1.01\pm0.06$  & $32.7\pm0.4$ & $10.4\pm0.2$
\\
\hline
 BH & 3.5 & 6 &   "  & no remnant  & $1.82\pm0.08$  & $3.67\pm0.12$ & $33.5\pm0.4$
\\
\hline
 BH & 4.0 & 6 &  $3.6 \times 10^{-3}$  & KINCUT=TRUE  & $0.037\pm0.004$  & $2.45\pm0.03$ & $0.089\pm0.006$
\\
\hline
 BH & 4.0 & 6 &   "  & KINCUT=FALSE  & $0.102\pm0.006$  & $3.02\pm0.03$ & $1.44\pm0.02$
\\
\hline
 BH & 4.0 & 6 &   "  & no remnant  & $0.198\pm0.008$  & $0.36\pm0.01$ & $3.18\pm0.03$
\\
\hline
$t\bar{t}$ & - & - & 469 & - & $0$ &   $0$   & $0.75\pm0.75$
\\
\hline
\end{tabular}
}
\caption{List of generated Monte Carlo samples.
Also reported are the expected numbers of events produced at the LHC
for an integrated luminosity of $1\,$fb$^{-1}$ at $\sqrt{s}=7\,$TeV and
$14\,$TeV after basic selection cuts on the $P_T$ of the leading lepton
(electron or muon), missing $P_T$ and visible transverse energy,
for different values of $M_{\rm G}$ and $D$.
The expected events for one of the most relevant SM process,
the $t\bar{t}$ production, are also reported.}
\label{tab:mcsamples}
}
\end{table}
\par
\subsection{Primary Emission}
\begin{figure}[tbh]
\centerline{\begin{tabular}{ r}
\centerline{\hbox{
\epsfig{figure=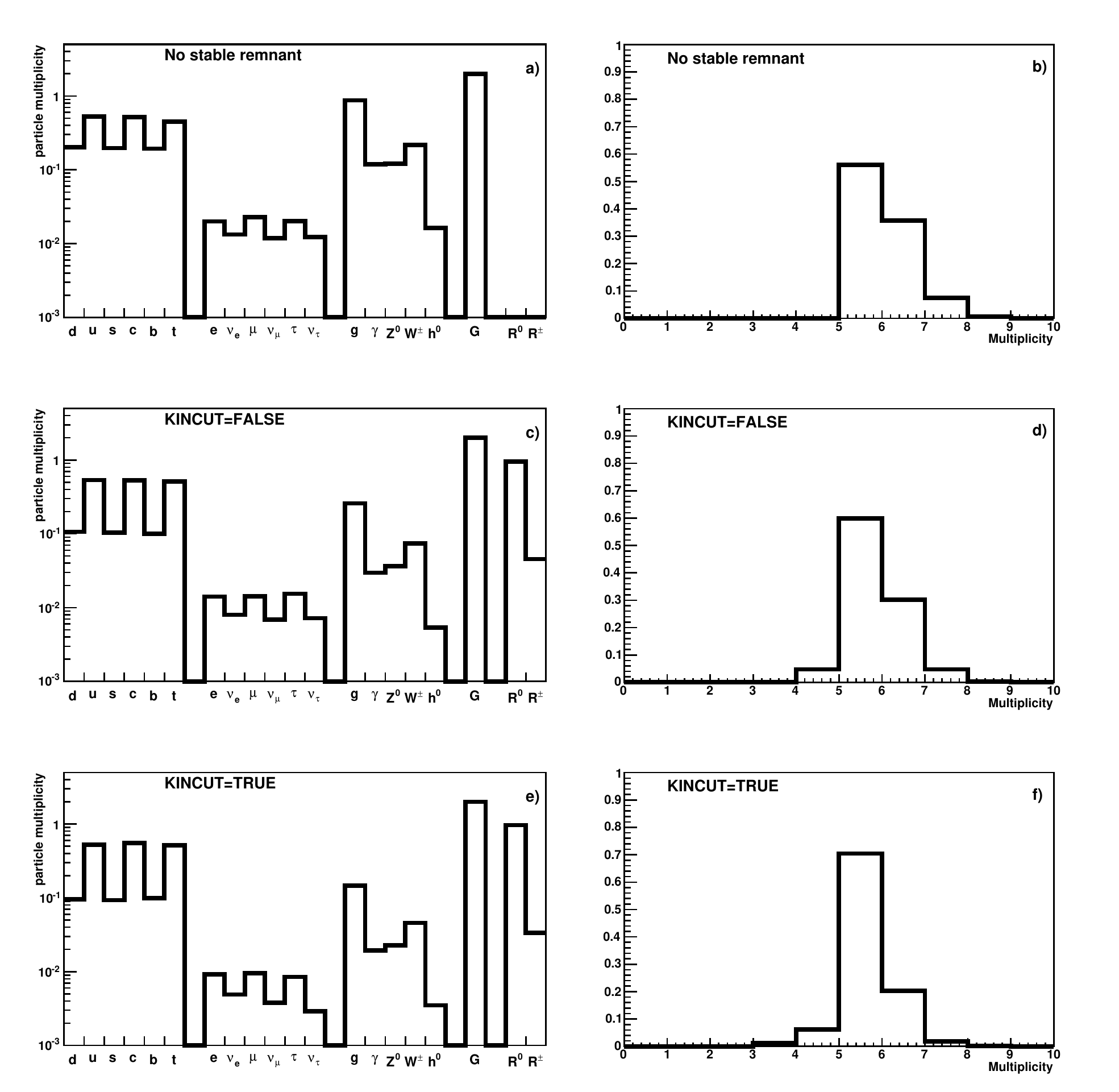,height=18cm,clip=} }}
\end{tabular}}
\caption{Primary decay products of the BH (panels a, c, e) and multiplicity
(panels b, d, f) for no remnant, KINCUT=FALSE and KINCUT=TRUE, respectively.
The plots have been obtained using Monte Carlo samples generated at $7\,$TeV,
with $M_{\rm G}=2\,$TeV in $D=8$.}
\label{fig:pid}
\end{figure}

Let us first examine the ``primary'' BH emission, namely the
particles produced by direct BH evaporation, before parton
evolution and hadronization are taken into account.
This emission is the direct output of CHARYBDIS2.
Fig.~\ref{fig:pid} shows the produced final state particles
and the particle multiplicity for $\sqrt{s}=7\,$TeV, $M_{\rm G}=2\,$TeV,
in $D=8$ and for the different remnant conditions.
The final state particles, listed in Fig.~\ref{fig:pid}, panels a), c) and e),
are grouped in SM particles (the first three blocks contain quarks, leptons 
and bosons, respectively), gravitons ($\rm G$), neutral ($\rm R^{0}$)
and charged ($\rm R^{\pm}$) BH remnants.
The remnants are neutral in the large majority of the cases,
with only a few percent of the events ending in a charged remnant.
The neutral remnants behave as weakly interacting massive particles (WIMPs),
producing a striking signal of missing transverse energy and a topology quite
different from the usual case of total BH evaporation. 
\subsection{Final Output}
\label{sec:finout} 
The primary output includes unstable particles, along with quarks and gluons that 
have to be processed by a parton shower Monte Carlo in order to produce the final state
observables (hadron level).
We used PYTHIA~\cite{pythia} to model the parton evolution
and hadronization of quarks and gluons emitted during the BH evaporation,
as well as to treat the unstable particle decays and the proton remnants.
This produced the final features of the events as would be seen by the LHC
detectors.
Besides the signal BH samples, also one of the most relevant SM
backgrounds for such topologies, the process $pp \to t\bar{t}X$, has been generated
using the NLO generator POWHEG~\cite{powheg}, interfaced with PYTHIA
for the parton shower, hadronization and particle decays.
At least one of the two top quarks was requested to decay in either $e$, $\mu$ or $\tau$.
\section{Results and outlook}
\setcounter{equation}{0}
\label{sec:results}
In order to point out the main characteristics of the different scenarios
under study and the possibility to observe a signal over the SM background, 
we have applied a basic selection to the hadron level Monte Carlo samples using
the following event variables:
\begin{itemize}
\item
$P_T^{\rm lep}$, defined as the transverse momentum of the leading lepton
($e$ or $\mu$) with $|\eta^{\rm lep}| < 2.5$, where $\eta^{\rm lep}$ is the lepton
pseudorapidity;
\item
the missing transverse momentum,
\be
P_T^{\rm miss}
=
\sqrt{\left(\sum_{i} P_{x_i}\right)^2 + \left(\sum_{i} P_{y_i}\right)^2}
\ ,
\ee
where $P_{x_i}$ and $P_{y_i}$ are the cartesian $x$ and $y$ components of the
momentum of the $i^{\rm th}$ particle, and $i$ runs over all the undetectable final
state particles:
neutrinos, gravitons and neutral BH remnant;
\item
the visible transverse energy,
\be 
E_T^{\rm vis}
=\sum_{k}\left(\sqrt{P_{x_k}^2+P_{y_k}^2}\right)
\ ,
\ee
where $k$ runs over all the detectable final state particles.
\end{itemize} 
The choice of the above simple variables is motivated by the fact that the
observation of final state particles with transverse momentum of several hundreds
of GeV's is the typical signal for the decay of a large mass state.
In particular, the requirement of high-energy leptons and/or high missing
transverse momentum allows to cope with the huge QCD background at the LHC.
Since BHs decay democratically to all SM particles, the search for extremely
energetic leptons is, in this context, one of the most direct way to look for deviations 
from the SM predictions. 
\par
\begin{figure}[tbh]
\centerline{\begin{tabular}{ r}
\centerline{\hbox{
\epsfig{figure=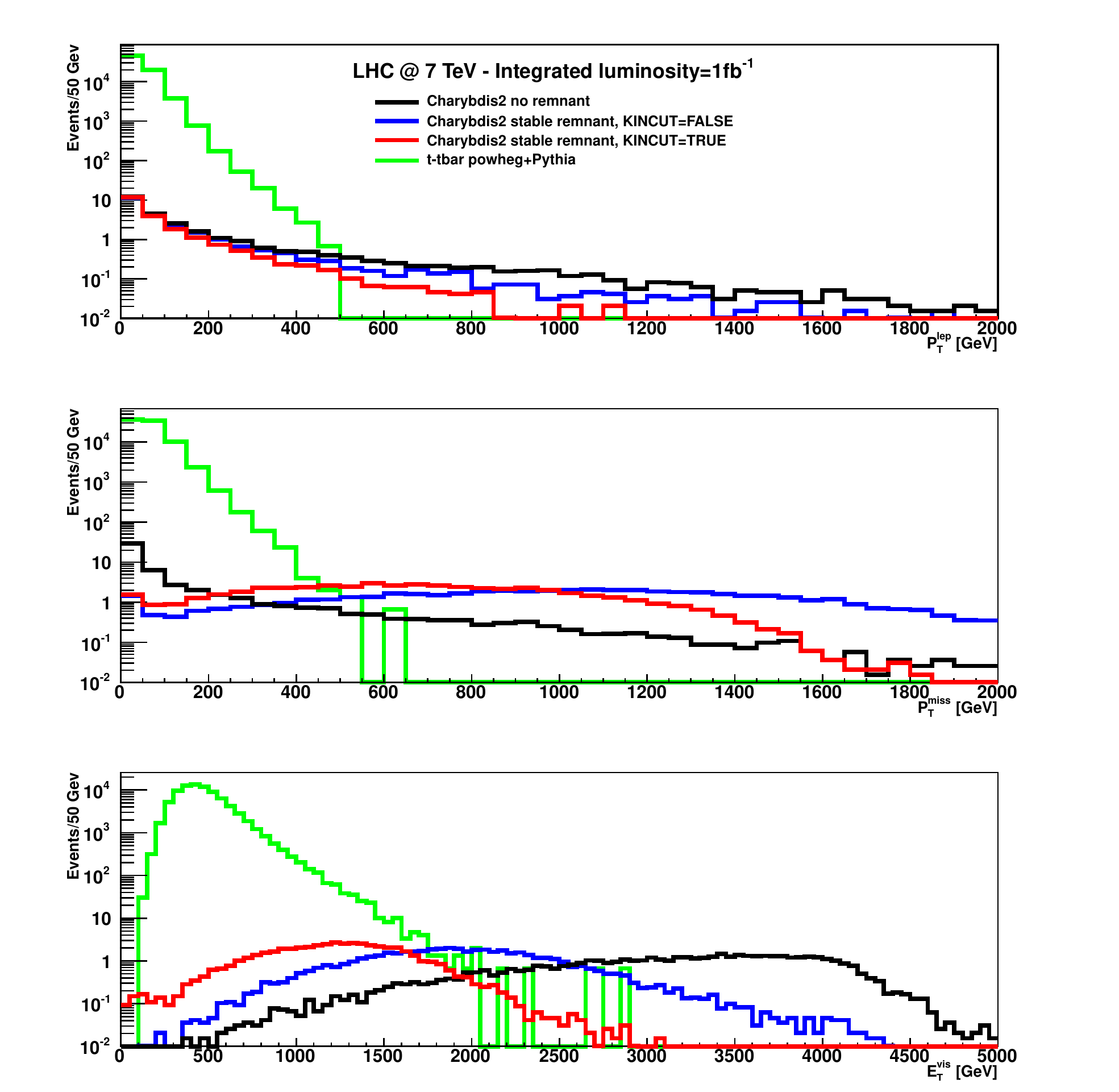,height=16cm,clip=} }}
\end{tabular}}
\caption{Distributions of $P_T^{\rm lep}$, $P_T^{\rm miss}$ and $E_T^{\rm vis}$ 
for BH events with different remnant scenarios and for the $t\bar{t}$ background.
The BH distributions have been obtained using the simulated samples at
$\sqrt{s}=7\,$TeV with $M_{\rm G}=2\,$TeV in $D=8$. 
}
\label{fig:dist_7tev}
\end{figure}
\begin{figure}[tbh]
\centerline{\begin{tabular}{ r}
\centerline{\hbox{
\epsfig{figure=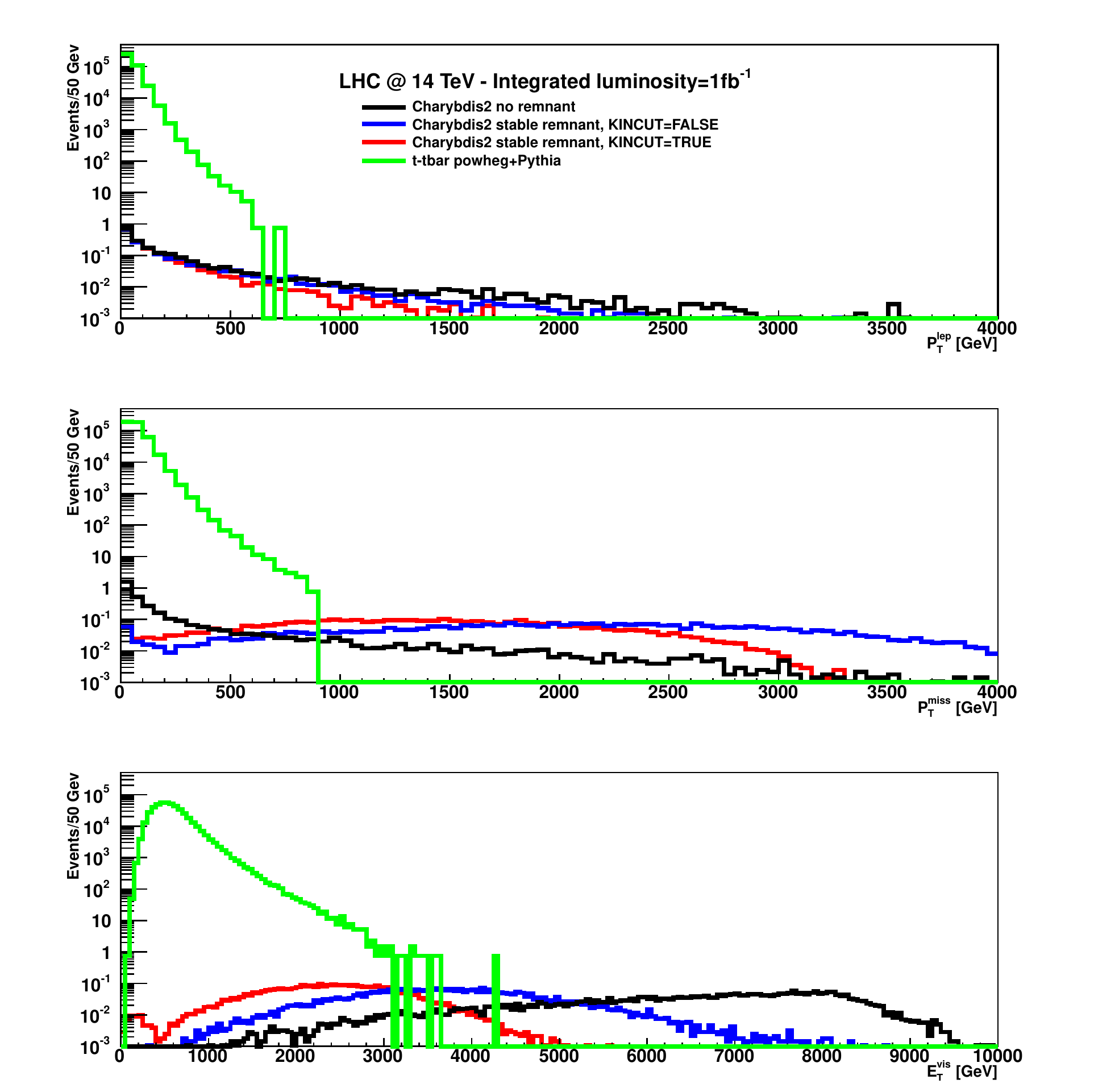,height=16cm,clip=} }}
\end{tabular}}
\caption{Distributions of $P_T^{\rm lep}$, $P_T^{\rm miss}$ and $E_T^{\rm vis}$ 
for BH events with different remnant scenarios and for the $t\bar{t}$ background.
The BH distributions have been obtained using the simulated samples at
$\sqrt{s}=14\,$TeV with $M_{\rm G}=4\,$TeV in $D=6$. 
}
\label{fig:dist_14tev}
\end{figure}

However, in case of stable neutral remnants, the lepton signal can be depressed
by the fact that a relevant fraction of the BH mass is not available in the decay.
On the other hand, the neutral remnant, behaving as a WIMP, will carry away a lot
of energy and enhance the missing transverse energy signal.
This is clearly visible in Figs.~\ref{fig:dist_7tev} and~\ref{fig:dist_14tev},
where we show, for two representative cases at $\sqrt{s}=7$ and $14\,$TeV,
the $P_T$ of the leading lepton, the $P_T^{\rm miss}$ and the $E_T^{\rm vis}$
for the different scenarios.
The $t\bar{t}$ background is also displayed. 
\par
\begin{figure}[tbh]
\centerline{\begin{tabular}{ r}
\centerline{\hbox{
\epsfig{figure=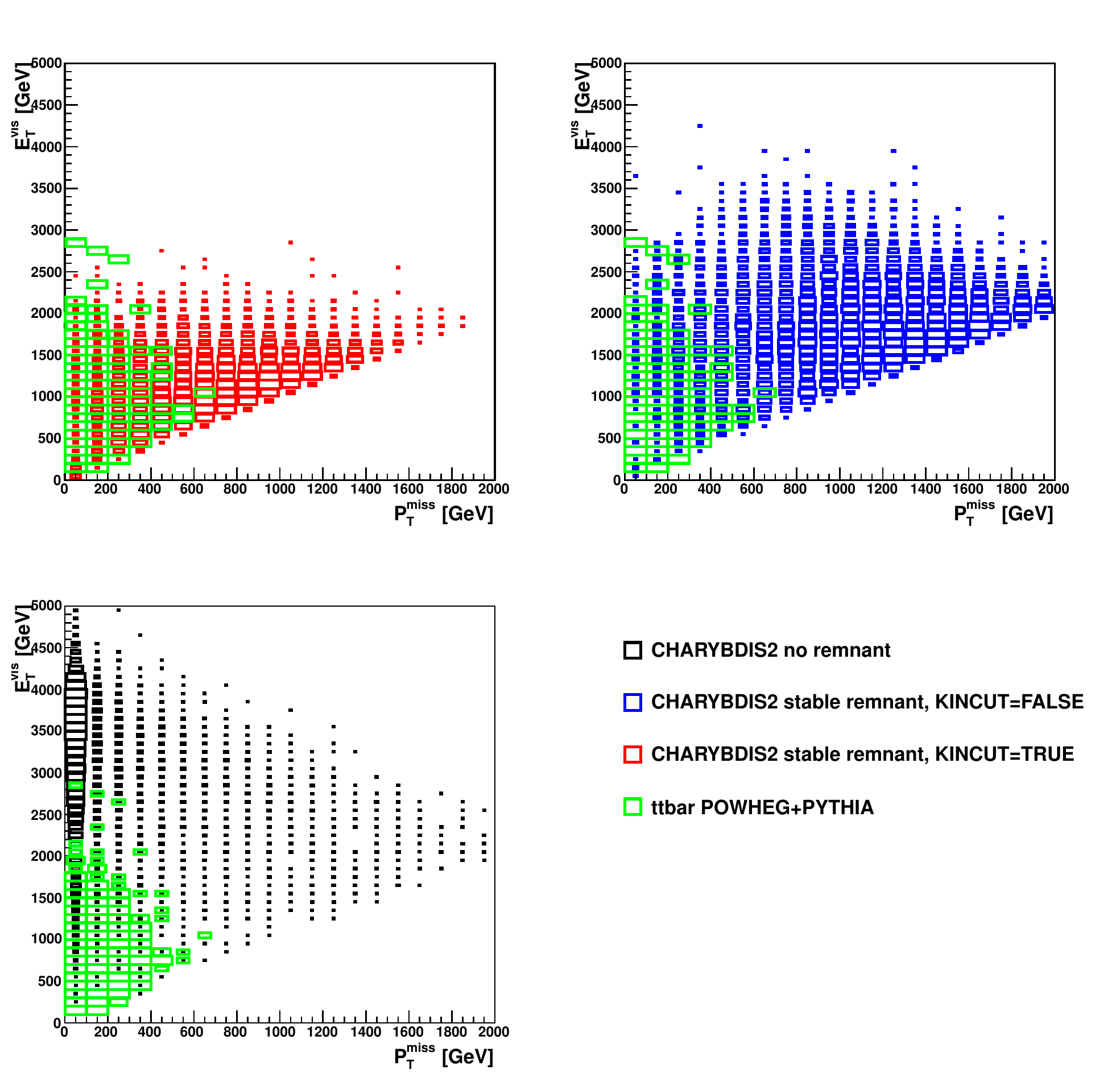,height=16cm,clip=} }}
\end{tabular}}
\caption{Distributions of signal and background events in the
$E_T^{\rm vis}$-$P_T^{\rm miss}$ plane for the different remnant scenarios. 
The BH distributions have been obtained using the simulated samples at
$\sqrt{s}=7\,$TeV with $M_{\rm G}=2\,$TeV in $D=8$. 
}
\label{fig:2dimdist_7tev}
\end{figure}

The main characteristics of the different cases are further clarified
in Fig.~\ref{fig:2dimdist_7tev}, which shows, for a representative case at
$\sqrt{s}=7\,$TeV, the event distribution in the $E_T^{\rm vis}$-$P_T^{\rm miss}$
plane for the different remnant scenarios and for the $t\bar{t}$ background.
If the BH decays completely, the events are concentrated at relatively low
$P_T^{\rm miss}$ and very high $E_T^{\rm vis}$, whereas, in the case of a stable
remnant, a shift towards much higher values of $P_T^{\rm miss}$ and consequently
lower $E_T^{\rm vis}$ is clearly visible. 
As already stressed in Section~\ref{sec:MCcode}, the control parameter KINCUT 
sets two extreme and probably not realistic descriptions of the stable remnant.
Nevertheless, given the large uncertainties in the theoretical picture,
these two borderline scenarios allow to quantify the possible spread
of the relevant experimental observables.
\par
Table~\ref{tab:mcsamples} summarizes the expected event yields,
after basic selection cuts on the event variables described above,
for the different BH scenarios and for the $t\bar{t}$ background~\footnote{In
the $t\bar{t}$ process, the high energy tails of the selection variables
are not well populated due to the exponential decrease of the production
rate with the energy.
Anyway, despite possible statistical fluctuations, 
the contribution of the background
in the signal region is very limited, 
given the very different high energy behaviour,
as is clearly shown in Figs.~\ref{fig:dist_7tev}
and~\ref{fig:dist_14tev}.}.
The reported values are normalized to an integrated luminosity of
$1\,$fb$^{-1}$ for $\sqrt{s}=7$ and $14\,$TeV.
The results show that stable remnant scenarios up to $M_{\rm G} \lesssim 2\,$TeV
can be already probed with the $5\,$fb$^{-1}$ of integrated luminosity
collected by the LHC during the last year of data taking.  
During the year 2012, we expect 4-to-5 times more integrated luminosity, which will
allow us to probe the fundamental mass in the range
$2\,$TeV\,$\lesssim M_{\rm G}\lesssim 2.5\,$TeV.
The sensitivity to higher values could be gained only by increasing the LHC energy,
since the production cross section for larger mass state is limited by the
parton density functions of the proton,
which steeply decrease going to larger momentum fraction of the partons. 
As shown in the second part of Table~\ref{tab:mcsamples}, $M_{\rm G} \simeq 4\,$TeV
could then be probed in the future, when the LHC runs at $\sqrt{s} = 14\,$TeV.
\par
Our findings should be mainly considered as an attempt to outline a reference picture, 
which can stimulate further investigations in such interesting stable BH remnant
scenarios.
Besides including specific theoretical models for BH remnants,
further developments should consider other possible SM backgrounds
(single-top, boson and di-boson production)
which, even if less important than $t\bar{t}$, can give non-negligible contributions.
The QCD production of events with two or more jets should also
be considered due to its huge cross section.
Since such processes can contribute to the background only for detector effects
leading to bad reconstruction of the experimental observables, a reliable estimation
of the expected background cannot disregard a detailed
detector simulation.
\section*{Acknowledgements}
We would like to thank the developers of CHARYBDIS2,
in particular M.O.P.~Sampaio and B.R.
Webber, for very useful suggestions and comments.
%
%\end{acknowledgements}
%
%
%
%
\end{document}